\documentclass{article}
\usepackage{placeins}

\usepackage{arxiv}

\usepackage[utf8]{inputenc} 
\usepackage[T1]{fontenc}    
\usepackage{hyperref}       
\usepackage{url}            
\usepackage{booktabs}       
\usepackage{amsfonts}       
\usepackage{nicefrac}       
\usepackage{microtype}      
\usepackage{cleveref}       
\usepackage{lipsum}         
\usepackage{graphicx}
\usepackage{natbib}
\usepackage{doi}

\title{\large Disentangling individual-level from location-based income uncovers socioeconomic preferential mobility and impacts segregation estimates}

\date{}

\newif\ifuniqueAffiliation
\uniqueAffiliationfalse

\ifuniqueAffiliation 

\author{ 
    \href{https://orcid.org/0009-0002-2555-095X}{\includegraphics[scale=0.06]{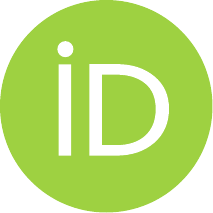}\hspace{1mm}} Marc Duran-Sala \\
    School of Civil and Environmental Engineering \\
    Ecole Polytechnique Fédérale de Lausanne
\And
    \href{https://orcid.org/0000-0003-0068-8047}{\includegraphics[scale=0.06]{orcid.pdf}\hspace{1mm}} Anandu Koikkalethu Balachandran\\
	School of Information\\
    University of Arizona
\And
    \href{https://orcid.org/0009-0005-0324-6094}{\includegraphics[scale=0.06]{orcid.pdf}\hspace{1mm}} Marta Morandini\\
	Department of Life and Health Sciences\\
    University of Aix-Marseille
\And
	\href{https://orcid.org/0000-0001-5085-0407}{\includegraphics[scale=0.06]{orcid.pdf}\hspace{1mm}} Timur Naushirvanov\\
	Department of Network and Data Science\\
    Central European University
\And
    \href{https://orcid.org/0000-0003-4415-4506}{\includegraphics[scale=0.06]{orcid.pdf}\hspace{1mm} Adarsh Prabhakaran }\\
	Department of Political Sciences  \\
    University College London
\And
	\href{https://orcid.org/0009-0007-9612-9464}{\includegraphics[scale=0.06]{orcid.pdf}\hspace{1mm}} Andrew Renninger\\
	Centre for Advanced Spatial Analysis \\
    University College London
\And
	\href{https://orcid.org/0000-0002-8756-5535}{\includegraphics[scale=0.06]{orcid.pdf}\hspace{1mm} Mattia Mazzoli} \\
	ISI Foundation
}
\else
\usepackage{authblk}

\newbox{\orcid}\sbox{\orcid}{\includegraphics[scale=0.06]{orcid.pdf}} 
\author[1]{%
	\href{https://orcid.org/0009-0002-2555-095X}{\usebox{\orcid}\hspace{1mm}Marc Duran-Sala}%
}

\author[2]{%
	\href{https://orcid.org/0000-0003-0068-8047}{\usebox{\orcid}\hspace{1mm}Anandu Koikkalethu Balachandran}%
}

\author[3]{%
	\href{https://orcid.org/0009-0005-0324-6094}{\usebox{\orcid}\hspace{1mm}Marta Morandini}%
}

\author[4]{%
	\href{https://orcid.org/0000-0001-5085-0407}{\usebox{\orcid}\hspace{1mm}Timur Naushirvanov}%
}

\author[5]{%
	\href{https://orcid.org/0000-0003-4415-4506}{\usebox{\orcid}\hspace{1mm}Adarsh Prabhakaran}%
}

\author[6]{%
	\href{https://orcid.org/0009-0007-9612-9464}{\usebox{\orcid}\hspace{1mm}Andrew Renninger}%
}

\author[7]{%
	\href{https://orcid.org/0000-0002-8756-5535}{\usebox{\orcid}\hspace{1mm}Mattia Mazzoli}%
}

\affil[1]{School of Civil and Environmental Engineering, Ecole Polytechnique Fédérale de Lausanne, Lausanne, Switzerland}
\affil[2]{School of Information, University of Arizona, Tucson, AZ, USA}
\affil[3]{Department of Life and Health Sciences, University of Aix-Marseille, Marseille, France}
\affil[4]{Department of Network and Data Science, Central European University, Vienna, Austria}
\affil[5]{Department of Political Sciences, University College London, London, UK}
\affil[6]{Centre for Advanced Spatial Analysis
University College London, London, UK}
\affil[7]{ISI Foundation, Turin, Italy}
\fi

\hypersetup{
pdftitle={The role of age, gender and income gaps on human mobility segregation},
pdfsubject={q-bio.NC, q-bio.QM},
pdfauthor={Mattia Mazzoli},
pdfkeywords={Mobility, Network, Segregation},
}

\begin{document}
\maketitle

\begin{abstract}


Segregation encodes information about society, such as social cohesion, mixing, and inequality. 
However, most past and current studies tackled socioeconomic (SE) segregation by analyzing static aggregated mobility networks, often without considering further individual features beyond income and, most importantly, without distinguishing individual-level from location-based income. 
Accessing individual-level income may help mapping macroscopic behavior into more granular mobility patterns, hence impacting segregation estimates.
Here we combine a mobile phone dataset of daily mobility flows across Spanish districts stratified and adjusted by age, gender and income with census data of districts median income. 
We build mobility-based SE assortativity matrices for multiple demographics and observe mobility patterns of three income groups with respect to location-based SE classes. 
We find that SE assortativity differs when isolating the mobility of specific income groups: we observe that groups prefer to visit areas with higher average income than their own, which we call preferential mobility. 
Our analysis suggests substantial differences between weekdays and weekends SE assortativity by age class, with weekends characterized by higher SE assortativity. 
Our modeling approach shows that the radiation model, which typically performs best at reproducing inter-municipal population mobility, best fits middle income and middle-aged flows, while performing worse on young and low income groups. 
Our double-sided approach, focusing on assortativity patterns and mobility modeling, suggests that state of the art mobility models fail at capturing preferential mobility behavior. 
Overall, our work indicates that mobility models considering the interplay of SE preferential behavior, age and gender gaps may sensibly improve the state of the art models performance.





\end{abstract}


\section*{Introduction}



Spatial and social segregation in living spaces has been shown to have a significant impact on daily life of residents \cite{massey1988dimensions}. 
Such segregation is often based on factors such as income, gender, ethnicity and age, and can dictate where people live, work and conduct their day-to-day activities \cite{schelling1969models,Zhang2019May}. 
Social factors, in turn, can directly and indirectly lead to affect access to healthcare and education, emphasizing social, economic, political and health disparities \cite{Hu2022Jan, Li2022Dec}. 

The study of segregation in living spaces has evolved significantly over the years, incorporating various aspects of human mobility. 
Traditional approaches focused mainly on residential segregation, assigning socioeconomic characteristics to individuals based on their home locations \cite{Liao2024Mar}. 
However, recent research has highlighted that mobility patterns play a crucial role for a more comprehensive understanding of spatial and social segregation \cite{Moro2021Jul, athey2021estimating}. 
Studies have demonstrated that incorporating social aspects such as income, age, gender and ethnicity into mobility patterns analysis can either reduce or amplify spatial and social segregation estimates \cite{ Moro2021Jul, Liao2024Mar,kazmina2024socio}. For instance, income segregation has been associated with differences in place and social exploration \cite{Moro2021Jul}. Moreover, mobility segregation estimates may change whether integration is measured on aggregate or individual levels, wealthy groups tend to travel longer distances \cite{farber2015measuring,barbosa2021uncovering,li2022aggravated}, while less affluent groups tend to make shorter trips \cite{wu2022human}.  
Gender also plays a significant role in mobility patterns, with women traveling more frequently, for shorter distances with multiple stop points and preferring public transport compared to men \cite{Law1999Dec,Acker2018Feb,Gauvin2020Jun}. 
As noted by \cite{Cresswell2016Apr}, "how people move (where, how fast, how often) is demonstratively gendered". 
Similarly, mobility patterns differ across age groups, beyond gender and income \cite{lenormand2015influence}, yielding a further level of complexity when it comes to intersectional groups. 
Moreover, women with lower socioeconomic status tend to walk more while women with higher socioeconomic status, education tend to commute using bicycle \cite{Yuan2023Mar}. 




Despite these advances in understanding segregation, several challenges remain in capturing the complexity of the interplay of demographic groups mobility patterns and segregation.
\begin{enumerate}
    \item Dynamic nature of segregation:  Many studies focus on static networks, overlooking potential differences that may occur on a daily scale, such as between weekdays and weekends \cite{nilforoshan2023human}.
    \item Interplay of individual and local income: most studies assign individuals' SE statuses based on their home-place income, however the interplay of individual income with location income may play an important role on mobility patterns and on SE segregation.
    \item Multidimensional segregation: The interplay between various demographic factors in shaping mobility patterns is not fully understood \cite{Bokanyi2021Oct}.
    \item Limitations of current human mobility models: population mobility models typically overlook the behavior of demographic groups and the role of different mobility purposes in shaping travel patterns \cite{zhao2023revisiting}.
\end{enumerate}


To address these challenges, here we leverage a public dataset of mobile phone records released by the Spanish Ministry of Transport, Mobility and Urban Agenda of Spain, \textit{MITMA}\cite{mitma}, encoding daily scale trips across Spanish districts stratified by age, gender and income of individuals, and study how segregation varies across demographic groups day by day. 
We cross these data with official data on median income provided by the Spanish National Statistical Office \cite{income_ine}, to map population mobility into ten location-based SE classes. 
We analyze how individuals of a given income class behave with respect to their home location in visitation patterns from home to destinations classified by mobility purpose (work/study, frequent and non-frequent destinations). 
Mobile phone data represent the most common data employed to study human mobility, bringing benefits to a wide range of disciplines ranging from transportation planning to public health \cite{murray2020digital,badr2020association,xiong2020mobile,grantz2020use,oliver2020mobile}. 
In transportation, these data are typically employed in the analysis of travel patterns and mobility flows \cite{xu2022planning}, traffic congestion and network performances \cite{essadeq2021use} and impact of emergencies and events \cite{escap2022using}, thus helping in effective transportation planning, traffic management and emergency responses to name a few. 
In public health, mobile phone data help in identifying and tracking the drivers of transmission \cite{murray2020digital}, estimating the exposure to environmental hazards \cite{hatchett2021mobility}, aiding contact tracing \cite{ming2020mobile,jahnel2020contact}, assessing the risk of virus importation from different regions \cite{bajardi2011human,balcan2009multiscale,pullano2020novel}, and informing public health policies \cite{grantz2020use}.

We uncover a further level of complexity in the way demographic groups behave in terms of mobility with respect to their residential SE class. To quantify this, we define a new measure of preferential mobility, that quantifies the behavior of income groups, defined on the individual level, with respect to their residential districts SE class.
Finally we dive into mobility modeling and assess what is the level of agreement of a common state of the art model, i.e. the radiation model \cite{simini2012universal}, at capturing mobility patterns of the various demographic groups. 
We find that the best agreement between modeled and observed trips is with middle income and middle-age groups, highlighting how this population mobility model does non perform well at reproducing mobility of low income, young age class demographic groups, especially for out-of-routine mobility. 


Our results have broad implications for human mobility, since current models and analyses that are based on aggregated origin-destination matrices may be missing important heterogeneities across demographic groups, with state of the art models performing better on certain classes, ages, and trips purposes and segregation estimates not considering both individual-level and location-based SE status.

\section*{Methods}

\subsection*{Data}
We use mobile phone data collected by a national mobile network operator in Spain \cite{ponce2021covid}, and published by the Ministry of Transport, Mobility and Urban Agenda of Spain, \textit{MITMA} in a public online repository \cite{mitma}. 
The data describe the hourly movements of individuals between Spanish districts from January 2022 to May 2024. 
Original districts defined by Spanish National Statistical Office (INE) are mapped into a coarser spatial division in which small districts with low population density are grouped to include areas not covered by antennas \cite{ponce2021covid}, resulting in $3792$ districts. 
The trips were aggregated using users' movements between consecutive \textit{stays} of at least 20 minutes in the same area, disregarding trips of less than 500 meters \cite{ponce2021covid}.
The data is aggregated in terms of origin-destination (OD) matrices at hourly time scale, encoding trips occurred during a given hour between two districts.  
Individuals belong to a given age-range (e.g. $0-25$, $25-45$, $45-65$, $65-100$ years), gender (e.g. female, male) and income (e.g. $<10$, $10-15$, $>15$ thousands euros per year) class. 
For some routes, users' features have been anonymized to ensure privacy. 
For each origin and destination, the activities at origin and destination are classified as home, work/study place, frequently and infrequently visited place.
This data collection is based on individuals' active events, e.g., users' calls together with passive events, in which the user's device position is registered due to changes in the cell tower of connection.
For the districts level distribution of income, we rely on the median income by consumption unit released by the Spanish National Statistical Office \cite{income_ine}.

\subsection*{Assortativity}
At the socioeconomic (SE) level of aggregation, we defined \textit{assortativity matrices} $X$ between income deciles $D$ for each SE class, following the approach in \cite{Bokanyi2021Oct}.
First, we used the probability of travelers living in districts of a given income decile $D = i$ to travel to districts with income decile $D = j$, the probability $C_{ij}$ of traveling from $i$ to $j$ is encoded in the \textit{probability matrix} as: 
\[
C_{ij} = \frac{
\sum_{ \{u  | D_{u, home = i}, D_{u, destination = j}\}} 1
}{
\sum_{ \{u  | D_{u, home = i}\}}  1
}
\]
where $u$ are the travelers residing in districts of decile $i$ and traveling to districts of $j$. Destinations can be stratified by trip purposes, defining a specific matrix encoding only trips with a given purpose. Given the normalized assortativity matrices $\tilde{X}$, where trips between $i$ and $j$ are normalized over the total trips occurring in the system, we compute the \textit{assortativity} $\rho$ with the Pearson correlation coefficient of the matrix entries, across all income deciles. 
A completely assortative matrix will have assortativity of value $\rho = 1$.
\[\rho = \frac{\sum_{i,j}ij \tilde{X}_{ij} - \sum_{i,j}i\tilde{X}_{ij}\sum_{i,j} j \tilde{X}_{ij}}
{
\sqrt{\sum_{i,j} i^2 \tilde{X}_{ij} - (\sum_{i,j} i \tilde{X}_{ij})^2}
\sqrt{\sum_{i, j} j^2\tilde{X}_{ij} - (\Sigma_{i,j} j \tilde{X}_{ij})^2}
}\]

\subsection*{Socioeconomic Preferential Mobility Index}

We introduce the \textit{socioeconomic preferential mobility index} $R$ to better understand how individuals are moving between socioeconomic districts.
Here, by computing the probability of moving from one SE class to another, we can compare the amount of flows towards richer SE classes with respect to those bound towards poorer SE classes. We define $R$ as:
\[ R = \frac{S_{upper} - S_{lower}}{S_{upper} + S_{lower}},\]
where $S_{lower}$ is the sum of the elements in the lower triangular matrix, i.e. towards lower SE classes and $S_{upper}$ the sum of the elements in the upper triangular matrix, i.e. towards higher SE classes, both without considering the matrix diagonal.
The index will be $R=1$ if no trips occur towards lower SE classes and $R=-1$ if no trips occur towards higher SE classes, whereas $R=0$ if the two amounts of trips are equally balanced. 

\subsection*{Mobility modeling}
We model flows between districts using a \textit{radiation model}. 
This model captures complex patterns and performs better than traditional gravity models for long-distance commutes beyond urban areas \cite{simini2012universal, lenormand2016systematic}. 
The model is specified as  
\begin{equation}
T_{pq} = O_i \frac{1}{1 - \frac{m_p}{M}}\frac{m_p m_q}{(m_p + s_{pq})(m_p + m_q + s_{pq})},
\end{equation}
where $T_{pq}$ is the average number of travelers from location $p$ to $q$, $O_p$ is the number of trips originating in $p$, $m_p$ and $m_q$ are the number of opportunities (here represented by population) at the origin and destination respectively, $s_{pq}$ is the number of opportunities (i.e. population) within a circle of radius $r_{pq}$ centered at $p$ (excluding source and destination), and $M = \sum_p m_q$ is the total number of opportunities. 
The model assumes that travelers choose destinations based on the quality of opportunities, represented by a fitness value $z$ drawn from a distribution $P(z)$. A traveler selects the closest opportunity with a fitness exceeding their threshold, also drawn from $P(z)$.   

We also compare the observed mobility patterns with the results of the radiation model to understand the structural biases in geography of Spanish mobility. 

\section*{Results}

\subsection*{Spatial mobility}

\begin{figure}
    \centering
    \includegraphics[width=0.9\textwidth]{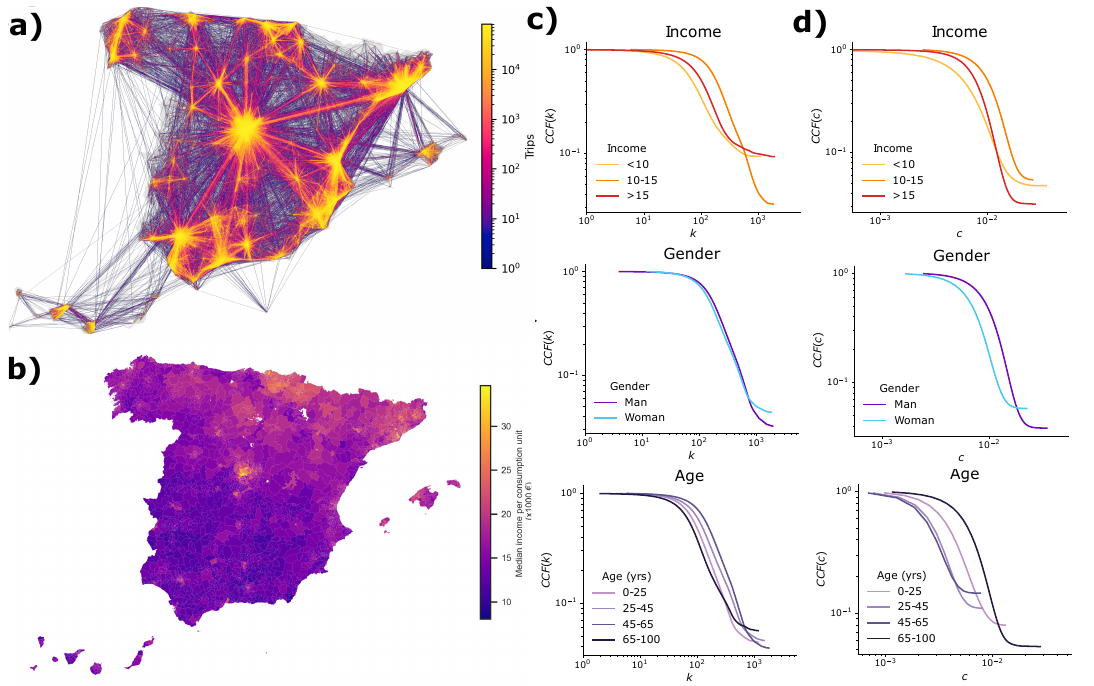}
    \caption{\textbf{Mobility and median income across Spain} \textbf{a)} Mobility network at Spanish districts level. Districts are represented as nodes, links are the total number of trips between them for the first week of September 2023, in log scale. \textbf{b)} Heatmap of median revenue per consumption unit of Spanish districts. \textbf{c)} Degree complementary cumulative density function (CCF) by income class, gender and age. The degree represents the number of unique districts visited by travelers of each district. \textbf{d)} The complementary cumulative density function (CCF) of the local clustering coefficient is analyzed by income class, gender, and age for each district.}
    \label{fig:panel1}
\end{figure}

Our analysis focuses on the spatial patterns of inter-district flows and economic indicators in Spain. Fig.\ref{fig:panel1} illustrates key findings from our study. \textbf{Inter-district flows}: Figure \ref{fig:panel1}a depicts the normalized amount of flows between Spanish districts on a logarithmic scale. The data represents trips recorded over a one-week period starting from the first of September, 2023.
\textbf{Economic disparities}: The median revenue by district is shown in Fig.\ref{fig:panel1}b. A clear north-south gradient is observable, with districts in the northern regions and in city centers generally exhibiting higher levels of wealth. This pattern indicates a significant level of economic disparity across the countries geography. \textbf{Spatial auto-correlation of income}: To quantify the spatial relationship of income distribution, we calculated the Moran's Index \cite{rey2009pysal,moran}. Our calculations returned a value of $0.73$ indicating a strong and significant spatial auto-correlation. This high value suggests that neighboring districts tend to have similar income levels, further reinforcing the observed geographical economic divide.

\paragraph{Network and spatial properties} 
To better understand the mobility patterns across income, gender and age we conducted a descriptive statistic analysis of the mobility network during the first week of September 2023.
The complementary cumulative function (CCF) based on degree $k$ and local clustering coefficient $c$ are shown in Fig.\ref{fig:panel1}c and d, respectively. The degree represents the number of unique districts visited by travelers from each home district, categorized by income, gender and age groups.

The degree distribution $CCF(k)$ reveals several key findings: (1) Middle-income travelers (10-15 thousand euros per year) exhibit the most extensive connectivity, visiting the widest range of unique places. This is evidenced by a slower and more gradual decline in their degree distribution. Low-income travelers (<10 thousand euros per year) show a sharp and early decline in their degree distribution, indicating that they visit fewer unique places, possibly due to financial constraints. High-income travelers (>15 thousand euros per year) exhibit a slightly delayed decrease compared to the lower-income group. This suggests that wealthier individuals tend to visit more unique places than low-income travelers, but not as many as the middle-income group. (2) Gender-group analysis reveals that men show a slightly later decline in $CCF(k)$ compared to women, indicating a marginally greater number of unique places visited by men. (3) Age-group mobility patterns show the highest degree of connectivity and largest range of places visited in the middle-age groups (25-45 and 45-64 years), as their degree distributions decline gradually and less steeply compared to other groups. The eldest (65-100 years) and the youngest (0-25 years) age groups exhibit sharper and faster declines in their degree distributions, indicating more constrained mobility patterns.




The local clustering distribution $CCF(k)$ weighted by the trips \cite{weightedclusteringcoefficient}, highlights the integration level of the network of visited districts for different traveler groups. This measure provides insights into how interconnected the travel patterns are for various demographic segments.  (1) High-income travelers demonstrate a more homogeneous pattern, indicating consistent mobility patterns that likely reflect well-established social structures in frequented locations. While the middle-income group exhibits more variability with a wider range of mobility patterns and social interactions. On the other hand, the low-income group shows a rapid decrease in the inter-connectedness of districts, reflecting limited mobility and lower levels of social integration. (2) Gender-group analysis reveals that on an average women exhibit a slower and more gradual decline in $CCF(k)$, indicating greater homogeneity in their mobility patterns. The overall inter-connectedness for women is lower, highlighting different social network structures and travel behaviors between genders. (3) Age-group mobility patterns indicate that the youngest (0-25 years) and eldest age group (65-100 years) show broader and more interconnected networks than the middle-aged groups (25-45 and 45-65 years), which can be related to less varied location visits and a more stable set of frequented places. 

These patterns further demonstrate that economic capabilities and social inclinations can give an intuition on the different pattern of mobility. 
\FloatBarrier

\begin{figure}[h]
    \centering
    \includegraphics[width=1\textwidth]{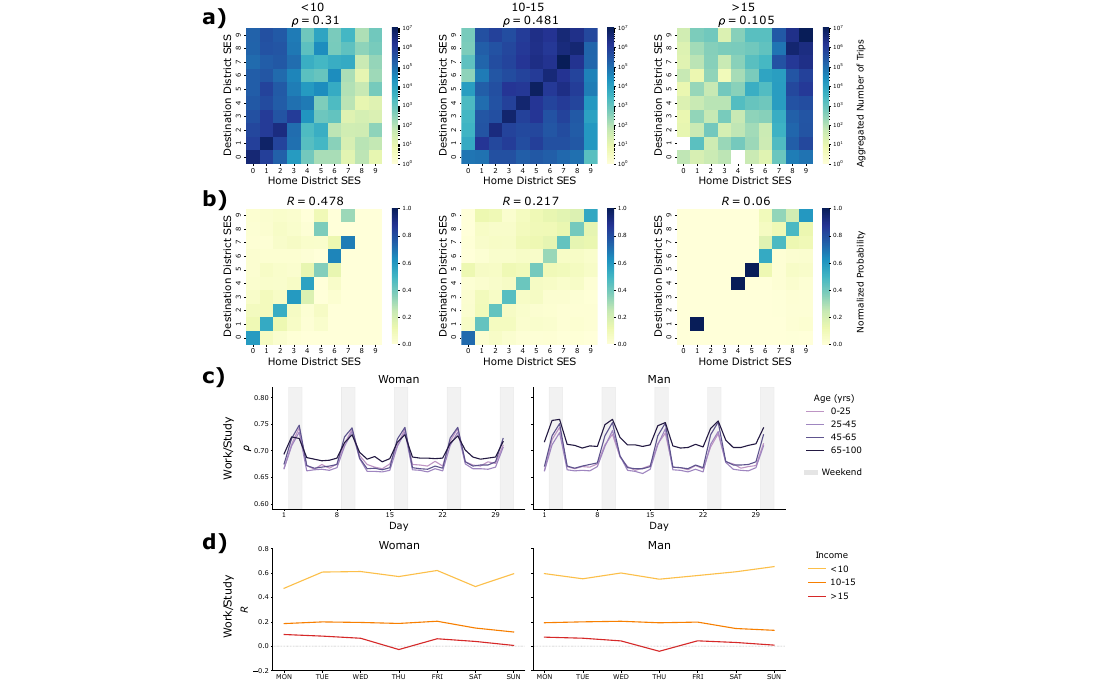}
     \caption{\textbf{Assortativity $\rho$ and Socioeconomic Preferential Mobility Index $R$} \textbf{a)} Assortativity matrices aggregated over a working week from 4-8 September 2023 across three income groups (<10, 10-15, >15 thousands euros per year). \textbf{b)} Probability matrices (normalized by Home District SES) aggregated over a working week from 4-8 September 2023 across three income groups (<10, 10-15, >15 thousands euros per year). \textbf{c)} Assortativity $\rho$ dynamics for travels from home to work or studying in September 2023 among different genderes and age groups (0-25, 25-45, 45-65, 65-100 years). \textbf{d)} Socioeconomic Mobility Preference Index $R$ aggregated by weekdays for travels from home to work or studying in September 2023 among different genderes and income groups (<10, 10-15, >15 thousands euros per year).}
    \label{fig:panel2}
\end{figure}

\subsection*{Assortativity and segregation analysis}
To uncover the level of segregation in mobility, we use two coefficients: we employ the known $\rho$ assortativity \cite{Bokanyi2021Oct} and introduce the $R$ Socioeconomic Mobility Preference Index. 
The assortativity being a measure of segregation across all the SE classes.
The socioeconomic preferential mobility index representing the overall tendency of segregation of the income group.

The assortativity matrices in Fig. \ref{fig:panel2}a illustrate the variations in travel patterns among the three income groups (<10, 10-15, >15 thousands euros per year). 
When considering the overall amount of trips, individuals from poorer classes frequently visit richer districts, while those from wealthier areas tend to travel often to poorer districts. 
The middle income group exhibits a more uniform distribution of mobility across middle-class districts, resulting in a notably higher assortativity coefficient. 
This indicates a greater correlation within the assortativity matrix and more segregated mobility patterns.

To examine potential mobility biases towards different income groups, we introduce a new metric, the Socioeconomic Preferential Mobility Index $R$, which quantifies the preference for moving from one's home to districts of higher or lower income levels for each income class. 
The probability matrices in Fig. \ref{fig:panel2}b demonstrate that, after normalizing mobility by the home district's socioeconomic status, all income groups show a tendency to travel more frequently towards the wealthiest districts. 
The mobility bias is most pronounced among the lowest income group and diminishes progressively with increasing income.

Fig. \ref{fig:panel2}c reveals the dynamics of assortativity $\rho$ in September 2023 across different genderes and age groups. 
The trend exhibits a weekly regularity, without systematic differences from one week to the next. Only weekends showing higher levels of assortativity compared to weekdays, meaning higher level of segregation during weekends. 
Additionally, there is a significant difference in assortativity levels among the oldest individuals, with men particularly displaying higher assortativity than women.

Fig. \ref{fig:panel2}d illustrates the dynamics of the Socioeconomic Preferential Mobility Index $R$ in September 2023 across different genderes and income groups, aggregated by weekday. 
The aggregation is done due to the weekly regularity of assortativity temporal trends, and to minimize noise due to multiple layers of stratification. 
Consistent with the patterns observed in the probability matrices in Fig\ref{fig:panel2}b, individuals from the lowest income group exhibit a consistently higher mobility preference towards the wealthiest districts. 
The same trend is present among middle-income groups, although to a lesser extent, and it diminishes further among the richest individuals. 
This effect remains consistent across all weekdays.

\begin{figure}[h]
    \centering
    \includegraphics[width=0.9\textwidth]{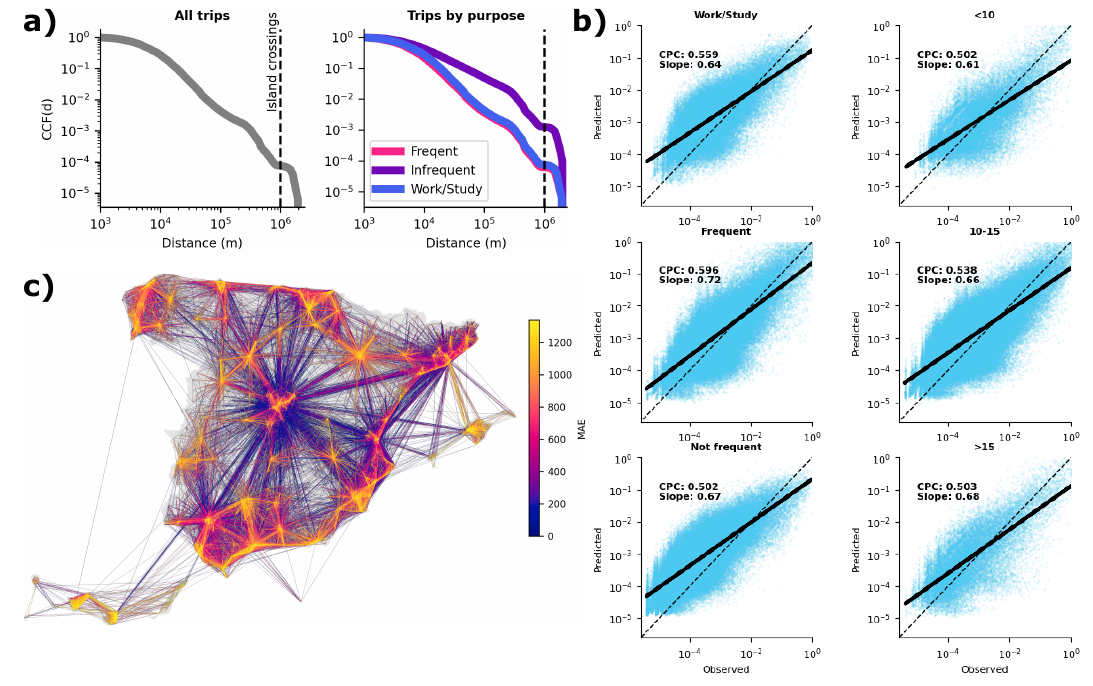}
    \caption{\textbf{Radiation model of mobility} \textbf{a)} Trip frequencies decay with distance, with a plateau representing trips that go further than the Iberian peninsula and thus cross to the islands. \textbf{b)} Comparison of mobility flows observed and predicted by the radiation model, stratified by mobility purpose and income group. CPC stands for the common part of commuters. \textbf{c)} Resulting mobility network capturing the mean absolute percentage error for all links in original network, clear colors representing big errors, darker colors representing small errors.}
    \label{fig:panel3}
\end{figure}

\subsection*{Radiation model simulation} 
Mobility decays with distance in the data: districts that are closer in space have a higher likelihood of trips between them. 
Fig. \ref{fig:panel3}a shows this decay for all trips and then disaggregated by mobility purpose, showing that infrequent trips typically reach farther distances than frequent trips and commutes. 
We see a kink in the cumulative distribution at a distance that corresponds to the width of the Iberian peninsula, suggesting that trips between mainland Spain and its islands exist off this decay function. 
This logic of distance decay allows us to implement the radiation model, which uses just population spatial distribution and distance to predict flows between areas. 

We use two measures to evaluate our model, common part of commuters and fitted slope.
We use the Sørenson-Dice index, also called Common Part of Commuters (CPC) in the study of mobility \cite{barbosa2018human}, to measure goodness-of-fit. 
This measure quantifies the similarity between the magnitude of observed and predicted flows and always lies within the interval [0, 1], with 1 indicating a perfect agreement between all predictions and observations and 0 indicating total disagreement at all links in the network. 
Our results are comparable to other models in a variety of contexts \cite{Lenormand2023Jan, simini2021deep, cabanas2023human}, despite fitting no free parameters and simulating a resolved spatial scale. 
We achieve a CPC above $0.5$ for all trip purposes and all socioeconomic classes. 
Our dataset enables us to understand how model performance varies for different groups and different kinds of trip, which should inform future work in mobility prediction.
Disaggregating the results in Fig. \ref{fig:panel3}b, the radiation model best approximates trips to frequented locations and trips by the middle income bucket. 
Surprisingly, the model does not perform best on work/study trips, since the radiation model is intended to model commuting and internal migration patterns rather than out-of-routine and frequent non-commuting day-to-day travel. 
This suggests that model evaluations employed in other work may be obscuring heterogeneous performance across different groups. 

Because of CPC's limitations, we also look at the slope, which tells us how close the radiation modelled travel probabilities are to the observed ones, with 1 being a perfect fit. Here we see again that frequent activities fall closest to this ideal but the results for income groups are mixed: the middle and high income groups are similar.

Not only is the model performance heterogeneous across trip characteristics, it is also spatially variable, which we see in Fig. \ref{fig:panel3}c: the model performs best predicting long journeys—especially between Madrid and peripheral cities, but it struggles to predict trips for low populated areas around Madrid and in  peripheral provinces.
As the distance decay would suggest, the model also does not perform well in the islands, where some of the highest error are. 
Spanish provinces around Madrid and in the coast are less populated so the model is also fitting flows between sparse and dense areas but erring with flows between sparse areas.  



\section*{Discussion}
This study highlights important issues in the study of human mobility and segregation estimates due to limitations such as income imputation from residential areas. By using a rich dataset of trips stratified by socioeconomic class, gender and age, we crossed travelers' individual-level income with their residential district median income and observed so far overlooked behaviors of preferential mobility between SE classes of locations. This allowed to uncover more granular income-related mobility patterns with respect to previous research, which potentially impacts segregation estimates.
We observed that individual-level income flows are more often bound towards higher income destinations. In order to quantify this, we introduced a measure of preferential mobility with respect to destination income, capturing the asymmetry between upper and lower triangles of the probability matrix. The preferential mobility index builds on studies of mobility assortativity \cite{napoli2023socioeconomic, Bokanyi2021Oct}, and sheds new light on the income directionality of mixing. This metric describes whether the mobility of population residing in given districts SE classes are systematically biased towards higher or lower destination SE classes, encoding the unbalance of mobility flowing predominately towards lower or higher deciles districts. We computed this metric separately for each individual-level income class, and observed that this behavior is more pronounced for travelers with low individual income, gradually diminishing for higher individual income groups.


We analyzed the mobility network of specific demographic groups and computed key network metrics like clustering coefficient and degree distribution, showing substantial differences, possibly reflecting economic capabilities and gender gaps that are in line with previous research \cite{barbosa2021uncovering,Gauvin2020Jun}.


From a modeling perspective, we showed here that trip purpose and individual features play an important role in characterizing aggregated origin-destination matrices. 
Our modeling approach, employing a radiation model, showed that the model performance varies across demographic groups, performing best for frequent, i.e. routine, mobility and middle income class trips. 
This suggests that current models may be unsuited for capturing mobility of specific groups, e.g. lowest income class, and to explain specific mobility purposes, like infrequent activities. 
We observed a sensibly different distance decay function for infrequent trips with respect to commutes and routine mobility, which may explain the scarce model performance on this type of trips. Contrary to our expectations, the radiation model does not perform as well for commutes as for frequent activities.  

Our work suggests that considering both individual and location-based income information when studying mobility patterns impacts socioeconomic segregation estimates and suggests that population mobility behaviors are not only explained by population spatial distribution and distance, but also by destinations income.
Moreover, adding SE information may help building new mobility models, helping increasing models performance, improving human mobility estimates in data desert regions, and better informing epidemic models for public health policies.



The dataset we used here contains rich stratification, and uses relatively high spatial resolution. 
However, for privacy reasons many minor routes that are characterized by a small number of trips were anonymized and we could not dive more into their demographic associations. 
For large cities, we were able to observe flows between districts within the metropolitan area, however for smaller towns we could only access flows bound to other towns or municipalities, especially in rural areas. 
Nevertheless, our results indicate that mobility mobility should incorporate information on individuals' income beyond their residential median income, typically used for individual income imputation, since these records exhibit heterogeneities that may tell a more complex story about the residents' mobility behavior. 
Further research should consider characterize the importance of this behavior for human mobility modeling at different scales.  

\section*{Data and code availability}
The data from Ministry of Transport, Mobility and Urban Agenda of Spain, MITMA ca be found here: \href{https://www.transportes.gob.es/ministerio/proyectos-singulares/estudios-de-movilidad-con-big-data/opendata-movilidad}{https://www.transportes.gob.es/ministerio/proyectos-singulares}.

The dashboard can be found here: \href{https://spain-mobility-dashboard.streamlit.app/}{https://spain-mobility-complexity-72h.streamlit.app/}.

The code can be found here: \href{https://github.com/adprabhak/Complexity72h_Mobility}{https://github.com/adprabhak/Complexity72h\_Mobility}.

\section*{Acknowledgments}
This work is the output of the Complexity72h workshop, held at the Universidad Carlos III de Madrid in Leganés, Spain, 24-28 June 2024. https://www.complexity72h.com

\nocite{*}
\bibliographystyle{abbrvnat}
\bibliography{references}  






\end{document}